# Photoionization-assisted, high-efficiency emission of dispersive wave in gas-filled hollow-core photonic crystal fibers

Yifei Chen,[1,2] Zhiyuan Huang,[1,2,5] Fei Yu,[3,4] Dakun Wu,[2,3] Jianhua Fu,[1,2] Ding Wang,[1] Meng Pang,[1,4] Yuxin Leng,[1,4,6] and Zhizhan Xu[1,7]

[1]*State Key Laboratory of High Field Laser Physics, Shanghai Institute of Optics and Fine Mechanics, Chinese Academy of Sciences, Shanghai 201800, China*
[2]*Center of Materials Science and Optoelectronics Engineering, University of Chinese Academy of Sciences, Beijing 100049, China*
[3]*R&D Center of High Power Laser Components, Shanghai Institute of Optics and Fine Mechanics, Chinese Academy of Sciences, Shanghai 201800, China*
[4]*Hangzhou Institute for Advanced Study, Chinese Academy of Sciences, Hangzhou 310024, China*
[5]*e-mail: huangzhiyuan@siom.ac.cn*
[6]*e-mail: lengyuxin@mail.siom.ac.cn*
[7]*e-mail: zzxu@mail.siom.ac.cn*

**We demonstrate that the phase-matched dispersive wave (DW) emission within the resonance band of a 25-cm-long gas-filled hollow-core photonic crystal fiber (HC-PCF) can be strongly enhanced by the photoionization effect of the pump pulse. In the experiments we observe that as the pulse energy increases, the pump pulse gradually shifts to shorter wavelengths due to soliton-plasma interactions. When the central wavelength of the blueshifting soliton is close to the resonance band of the HC-PCF, high-efficiency energy transfer from the pump light to the DW in the visible region can be obtained. During this DW emission process, we also observe that the spectral center of the DW gradually shifts to longer wavelengths leading to a slightly-increased DW bandwidth, which can be well explained as the consequence of phase-matched coupling between the pump pulse and the DW. In particular, at an input pulse energy of 6 μJ, the spectral ratio of the DW at the fiber output is measured to be as high as ~53% together with a conversion efficiency of ~19%. These experimental results, explained by numerical simulations, pave the way to high-brightness light sources based on high-efficiency frequency-upconversion processes in gas-filled HC-PCFs.**

## 1. INTRODUCTION

In the past decade broadband-guiding anti-resonant hollow-core photonic crystal fibers (HC-PCFs) filled with gases [1-3], have become excellent platforms for studying ultrafast nonlinear optics such as ultrashort pulse compression to the single-cycle regime [4], efficient generation of tunable DW at deep and vacuum ultraviolet (UV) wavelengths [5-8], soliton-plasma interactions [9-13], etc. These successful applications of the HC-PCF in nonlinear optics are generally due to the remarkable feature of the HC-PCF – it can tightly confine light in its μm-sized core while maintaining low-loss guidance over long distances [1], largely enhancing the light-matter interactions in it. In addition, when filled with gases, the HC-PCF can offer waveguide-induced anomalous dispersion across a broad wavelength range through tuning the gas pressure, leading to versatile soliton dynamics [2,3].

While the anti-resonant HC-PCF can provide low-loss optical guidance over a broad wavelength range, its transmission window is discontinuous, disrupted by the presence of several narrow-band resonances [14,15]. Within these resonance bands the guidance of light shows high loss due to the energy coupling between the core and the cladding modes in the fiber. Tani *et al.* demonstrated the influence of this transmission discontinuousness on some ultrafast nonlinear processes, such as the suppression of both the soliton self-compression and UV DW generation [16]. They also predicted that when the central wavelength of the pump light is close to the resonance band of the anti-resonant fiber, strong narrow-band emission of phase-matched DW could be obtained due to the sharp dispersion slope within the resonance band of the fiber [16]. Recently, Sollapur *et al.* experimentally demonstrated the generation of more-than-three-octave supercontinuum light from 200 nm to 1.7 μm that spans multiple resonant bands in a Kr-filled anti-resonant HC-PCF [17], in which the emission of multiple phase-matched DWs was involved. More recently, Meng *et al.* demonstrated in the experiments that near-infrared DW emission can be obtained in the resonance band of an Ar-filled anti-resonant fiber [18,19]. In this experiment, in order to achieve high-efficiency frequency conversion the central wavelength of the pump pulses was tuned to 1300 nm, close to the ~1000 nm resonance of the fiber being used [19]. All of these theoretical and experimental studies indicate that the high-efficiency DW generation can be achieved only if the wavelengths of the pump pulses are close to the resonances of the anti-resonant fibers. In contrast, the narrow-band DW emission can be minimized when the pump pulse is spectrally tuned far from the fiber resonance.

In our recent experiments, we observed, however, that the frequency blueshift of the plasma-driven soliton, in a short length He-filled anti-resonant fiber, could be disrupted due to the DW emission at the fiber resonance wavelength [11], even though the pump wavelength (at 800 nm) was quite far away from the fiber resonance at ~470 nm. Soliton self-compression results in the fast accumulation of plasma, leading to a strong spectral blueshift of the 800 nm pump pulse and therefore triggering the emission of narrow-band DW.

In this work, we further enhance this photoionization-assisted, high-efficiency DW generation process through using a 25-cm-long He-filled singe-ring photonic crystal fiber (SR-PCF) with a fiber resonance at ~553 nm. In the experiments, we observed that as the energy of pump pulse increased, the strong soliton-plasmas interaction rapidly shifted the pump pulse to shorter wavelengths. When the pedestal of the pulse spectrum approached one resonance band of the fiber, high-efficiency energy transfer to the narrow-band DW can be triggered. As the pump pulse prorogated further in the gas-filled fiber, the DW exhibits an increased spectral width as well as a significant spectral redshift, both of which can be well explained using the phase-matched condition of the soliton-DW coupling. When the energy of the pump pulse was 6 μJ, we obtained, at the fiber output, a notable DW spectral ratio as high as ~53% together with a high conversion efficiency of ~19%. These studies, on one hand, point out the fact that assisted by the soliton blueshift effect due to photoionization, the strong phase-matched DW emission could still happen, even though the wavelength of the pump pulse is tuned to be far from the fiber resonances. On the other hand, the combination of the two frequency-upconversion mechanisms, paves the way to high-efficiency light source generation in gas-filled anti-resonant HC-PCFs.

## 2. EXPERIMENTAL RESULTS
### A. Experimental Set-up

The experimental layout is sketched in Fig. 1(a). A Ti:Sapphire laser system, which delivers ~45-fs pulses with an energy of ~0.3 mJ at 1-kHz repetition rate, was used to pump a 1-m-long Ar-filled hollow-core fiber (HCF). The HCF with 250-μm core diameter was put in a gas cell sealed by 0.5-mm-thick fused silica (FS) windows and filled with 212-mbar Ar gas. Due to self-phase modulation (SPM) effect, the pump pulses were spectrally broadened during propagation in this Ar-filled HCF. Chirped mirrors (CMs) were used to compress the pulses to <20 fs. A half-wave plate (HWP) and a wire grid polarizer (WGP) were used to control the pulse energy launched into the SR-PCF. The compressed pulses were coupled into a 25-cm-long He-filled SR-PCF with core diameter of 24 μm and gas pressure of 10.9 bar using a coated lens with a focal length of 10 cm. Both ends of the gas cell were sealed by 1.5-mm-thick coated FS windows. The spectra at the output of the SR-PCF were measured by a fiber spectrometer (Ocean Optics HR2000+) connected with an integrating sphere.

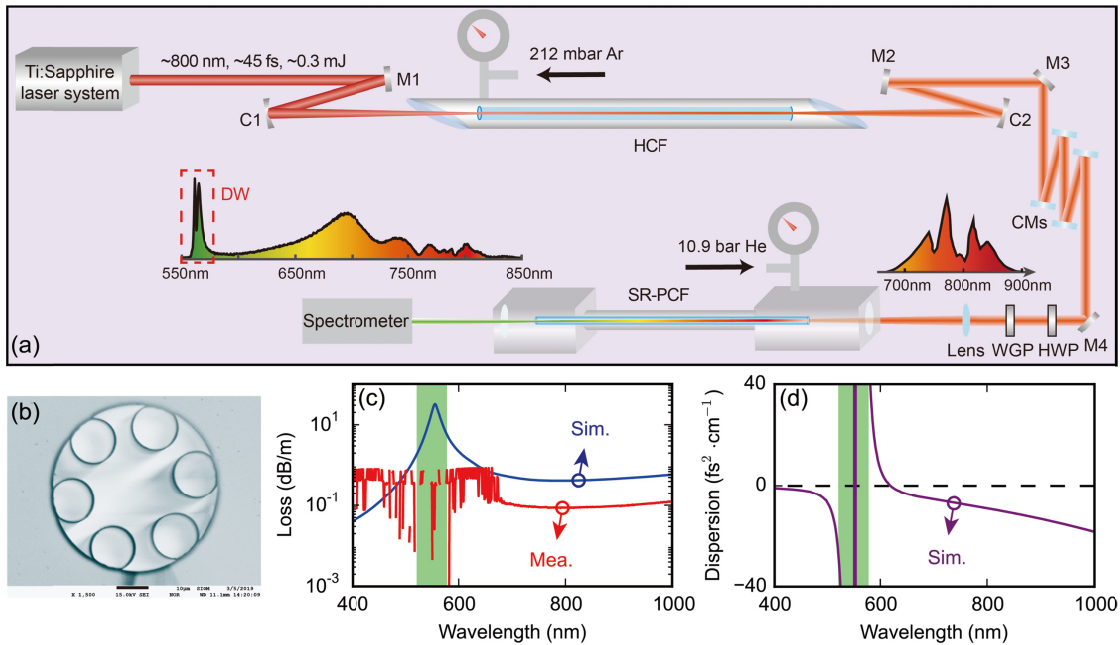

**Fig. 1.** (a) Schematic of the experimental set-up, including a stage for nonlinear pulse compression and a stage for DW generation. M1-M4, silver mirrors; C1-C2, concave mirrors; CMs, chirped mirrors; HWP, half-wave plate; WGP, wire grid polarizer. These two insets in panel (a) represent the spectra of the input and output pulses in the He-filled SR-PCF at the input pulse energy of 3.7 μJ. (b) SEM of the SR-PCF with 24-μm core diameter and ~0.26-μm capillary wall thickness used in the second stage. (c) Measured (red solid line) and simulated (blue solid line) fiber losses of the fundamental mode $HE_{11}$ of the SR-PCF. The simulated fiber loss was calculated by the BR model. (d) Simulated dispersion (purple solid line) of the SR-PCF filled with 10.9-bar He gas, calculated by the ZS model. In both (c) and (d), the light green bars indicate the first resonant spectral region of the SR-PCF.

Figure 1(b) shows the scanning electron micrograph (SEM) of the SR-PCF. The SR-PCF consisting of six thin-walled capillaries arranged around a hollow core has a wall thickness ~0.26 μm, leading to the first resonant wavelength of ~553 nm [20,21]. In Fig. 1(c), we plot the measured (red solid line) and simulated (blue solid line) fiber losses of the fundamental model of the SR-PCF. The simulated fiber loss was calculated by the bouncing ray (BR) model [22]. Figure 1(d) shows the simulated dispersion of the SR-PCF at 10.9-bar He gas based on a fully analytical model developed by Zeisberger and Schmidt (ZS), which can describe the dispersion in the resonant and anti-resonant spectral regions [23]. The light green bars in Figs. 1(c) and 1(d) point out the first resonant spectral region of the SR-PCF, where the dispersion curve shows a sharp change due to the core-cladding resonance.

**B. High-Efficiency DW Generation**

In Fig. 2(a), we measured the spectral evolutions at the output of a 25-cm-long He-filled SR-PCF as a function of input pulse energy from 1 μJ to 6 μJ, and the normalized spectral intensities (on the linear scale) for different input pulse energies [corresponding to the positions marked as black dotted lines in Fig. 2(a)] are shown in Fig. 2(b). At the input pulse energy of <2 μJ, the pulses mainly show a SPM-induced spectral broadening. While the input pulse energies used in the experiments are set at the range between 2 μJ and 3 μJ, the generated plasma results in a spectral blueshift. Meanwhile, the spectra also exhibit some spectral peaks near the first resonant wavelength of ~553 nm, which can be understood as a consequence of phase-matched DW generation.

In particular, for input pulse energies of >3 μJ, we observe that the pulses present a clean spectral blueshift and the central wavelength of the plasma-driven blueshifting soliton [marked as (i)] continuously moves to the shorter wavelengths as input pulse energy increases. As shown in Fig. 2(c), the soliton wavelength (blue circle line) defined as the spectral centroid of the blueshifting soliton can be tuned to ~660 nm at the input pulse energy of 6 μJ. The interesting thing is that the resonance-induced phase-matched DW [marked as (ii)] is also enhanced due to continuously spectral blueshift when gradually increasing input pulse energy. Simultaneously, the generated DW shows not only an increased spectral width, but also an enhanced spectral redshift. In Fig. 2(c), the DW wavelength (red square line), also defined as the spectral centroid of the DW, can be continuously tuned from ~560 nm to ~572 nm through adjusting input pulse energy. In addition, as shown in Fig. 2(d), the conversion efficiency (green circle line) defined as the ratio of the DW energy to the input pulse energy and the DW spectral ratio (orange square line) of the DW to the total output spectrum both increase with input pulse energy. At the input pulse energy of 6 μJ, the DW spectral ratio is ~53% and its spectral width [full width at half maximum] is broadened to ~15 nm, together with a high energy conversion efficiency of ~19%.

This resonance-induced high-efficiency DW generation is a direct consequence of phase matching between the plasma-driven blueshifting soliton and the DW. The blueshifting soliton is spectrally close to the resonant spectral region of the SR-PCF, which enhances the energy transfer from the blueshifting soliton to the DW, leading to a high-efficiency frequency upconversion process. We will explain these experimental observations in the section of numerical analysis. It should be noted that the conversion efficiency depends on the coupling efficiency and can be further improved through optimizing experimental conditions. For example, a pressure gradient can be used to decrease the energy loss at the input of the SR-PCF.

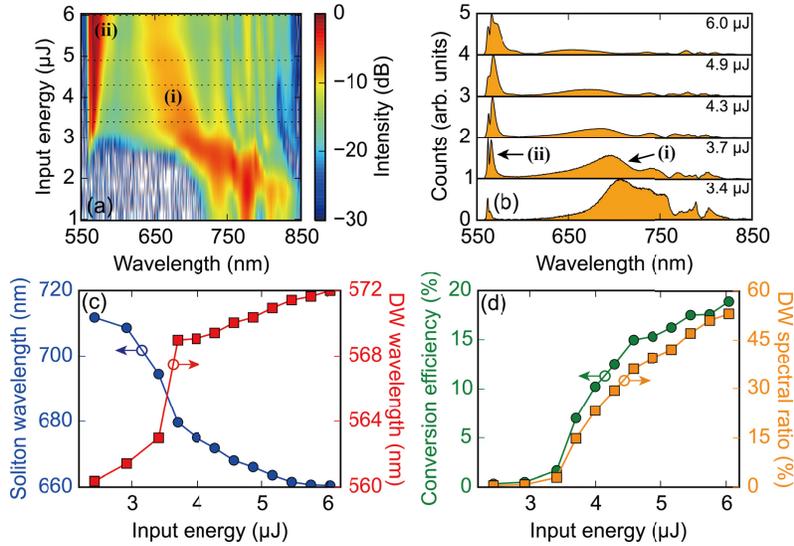

**Fig. 2.** (a) Measured spectral evolutions at the output of a 25-cm-long SR-PCF with 24-μm core diameter and 10.9-bar He gas as a function of input pulse energy. (b) The normalized spectral intensities (on the linear scale) for different input pulse energies [marked as black dotted lines in panel (a)]. In both (a) and (b), (i) and (ii) represent the blueshifting soliton and DW, respectively. (c) Soliton wavelength (blue circle line) and DW wavelength (red square line) as a function of input pulse energy. (d) Conversion efficiency (green circle line) and DW spectral ratio (orange square line) versus input pulse energy.

## 3. NUMERICAL ANALYSIS

To investigate the pulse propagation along the He-filled SR-PCF and the mechanism of the resonance-induced high-efficiency DW generation, we performed the numerical simulations using a well-known single-mode unidirectional field equation [24,25]. In the simulations, the fiber loss was calculated by the BR model and the dispersion was simulated by the ZS model, and the gas ionization was calculated by the Perelomov-Popov-Terent'ev model [26] modified with the Ammosov-Delone-Krainov coefficients [27]. In addition, the input pulses were characterized using a home-built second-harmonic generation frequency-resolved optical gating (SHG-FROG) and used in the simulations (the lens- and window-induced dispersion is also included in the pulses). As shown in Fig. 3(a), we plot the simulated spectral evolutions after propagating a 25-cm-long SR-PCF (24-μm core diameter and ~0.26 μm capillary wall thickness) filled with 10.9-bar He gas as a function of input pulse energy. When input pulse energies range from ~3 μJ to ~4.8 μJ, we can observe some resonance-induced narrow-band spectral peaks. Especially for pulse energies of >4.8 μJ, as the plasma-driven blueshifting soliton [marked as (i)] rapidly approaches to the resonant spectral region, the spectral width of the phase-matched DW [marked as (ii)] becomes wider and its spectra show an obvious redshift, which qualitatively agree with the experimental results in general.

In order to explain the resonance-induced intense DW generation, we show the simulated temporal and spectral evolutions as a function of position in the SR-PCF at an input pulse energy of 6.5 µJ [marked as black dotted line in Fig. 3(a)]. In the time domain, the input pulses in the anomalous dispersion region are gradually compressed to shorter duration at the first few centimeters due to soliton-effect self-compression. At the position of ~4.4 cm, the self-compressed pulses reach a maximum peak power of ~1.2 GW, generating a maximum plasma density of ~$8.2\times10^{16}$ cm$^{-3}$ [see white circle line in Fig. 3(c)]. Meanwhile, the fast accumulation of plasma results in a rapid spectral blueshift. The plasma-driven blueshifting soliton is regarded as a pump light tuned to shorter wavelengths and gradually approaches the resonant spectral region, leading to the emission and amplification of the phase-matched DW.

However, to explain the bandwidth broadening and the spectral redshift of the DW mentioned in both experiments and simulations, analysis on the phase-matched condition between the blueshifting soliton and the DW is required. The phase-matched condition can be expressed as [28,29]:

$$\Delta\beta(\omega_{DW}) = \beta_{DW}(\omega_{DW}) - \beta_{Sol}(\omega_{DW}) = 0, \quad (1)$$

with

$$\beta_{Sol}(\omega_{DW}) = \beta(\omega_{Sol}) + \beta_1(\omega_{DW} - \omega_{Sol}) + \beta_{Kerr} + \beta_{Plasma}, \quad (2)$$

where $\Delta\beta$ is the phase mismatch at frequency $\omega_{DW}$, $\beta_{DW}$ and $\beta_{Sol}$ are the propagation constants of the DW and the blueshifting soliton, $\beta(\omega_{Sol})$ and $\beta_1$ are the propagation constant and the inverse group velocity at the soliton central frequency $\omega_{Sol}$. $\beta_{Kerr} = \gamma P_{SC}\omega_{DW}/\omega_{Sol}$ and $\beta_{Plasma} = -\omega_{Sol}^2\rho/2n_0 c\rho_{cr}\omega_{DW}$ are the nonlinear contributions from the optical Kerr effect and the plasma, where $\gamma$ is the nonlinear parameter, $P_{SC}$ is the peak power of the self-compressed pump pulse, $\rho$ and $\rho_{cr}$ are the plasma density and the critical free-electron density, $n_0$ and $c$ are the linear refractive index at the $\omega_{Sol}$ and light speed. Here, we divide the phase mismatch into two parts:

$$\Delta\beta(\omega_{DW}) = \beta_A(\omega_{DW}, \omega_{Sol}) - \beta_B = 0, \quad (3)$$

with

$$\beta_A(\omega_{DW}, \omega_{Sol}) = \beta_{DW}(\omega_{DW}) - \beta(\omega_{Sol}) - \beta_1(\omega_{DW} - \omega_{Sol}), \quad (4)$$

and

$$\beta_B = \beta_{Kerr} + \beta_{Plasma}. \quad (5)$$

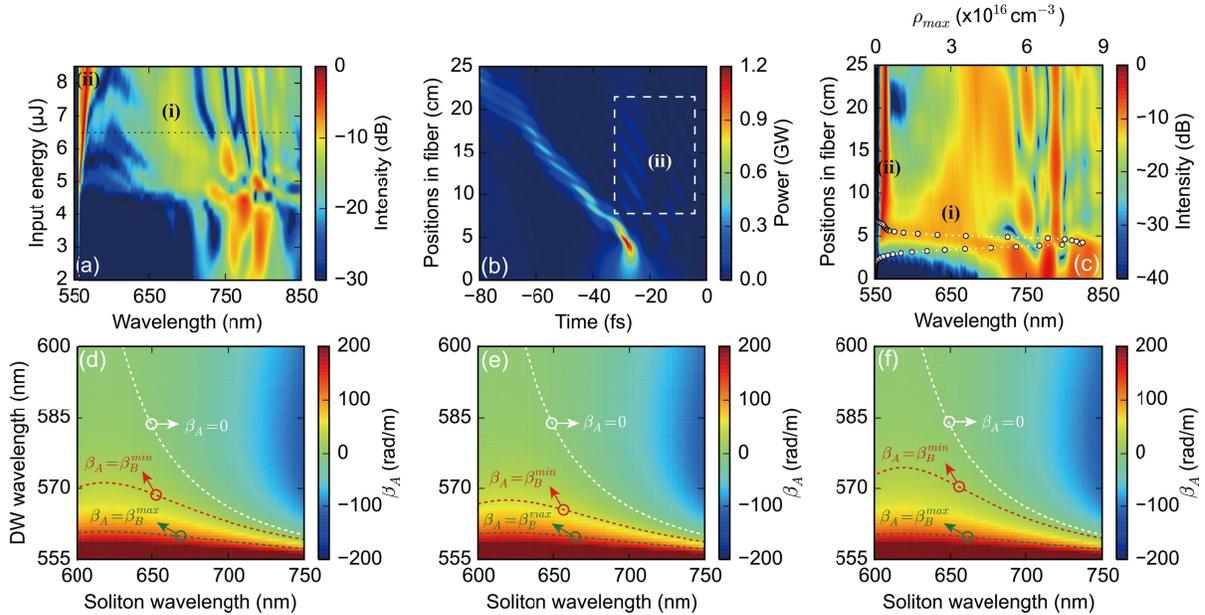

**Fig. 3.** (a) Simulated spectral evolutions at the output of a 25-cm-long SR-PCF as a function of input pulse energy. (b) and (c) Simulated temporal and spectral evolutions for different positions in fiber at the input pulse energy of 6.5 µJ [marked as black dotted line in panel (a)]. The white circle lines in panel (c) show the maximum plasma density. In simulations, the SR-PCF has a core diameter of 24 µm and a wall thickness of ~0.26 µm, and it was filled with 10.9-bar He gas. The pulses were characterized by SHG-FROG and used as the input. The fiber loss was simulated by BR model and the dispersion was calculated by ZS model. In (a)-(c), (i) and (ii) correspond to the blueshifting soliton and DW. (d)-(f) Phase-matched conditions between the blueshifting soliton and DW, the corresponding input pulse energies are 6.5 µJ, 5.5 µJ and 7.5 µJ, respectively.

In Fig. 3(d), we first plot the dispersion-related term $\beta_A$ with different soliton wavelength and DW wavelength, and the white dashed line indicates the condition of $\beta_A = 0$. Then we draw two contour lines with the maximum (green dashed line) and minimum (red dashed line) values of the $\beta_B$ in Fig. 3(d). It should be noted that we calculated the Kerr- and plasma-induced terms $\beta_B$ for every position in fiber at the input pulse energy of 6.5 µJ. Therefore, the phase-matched DWs are generated in the region between $\beta_B^{min}$ and $\beta_B^{max}$. Moreover, we can also observe the spectral width of the DW becomes wider and its spectrum broadens to longer wavelengths as the soliton wavelength increases, showing a spectral redshift. Furthermore, we also plot the phase-matched conditions when the input pulse energies are 5.5 µJ and 7.5 µJ, as shown in Figs. 3(e) and

3(f). For a lower pulse energy, we can see that the maximum spectral width and spectral redshift in Fig. 3(e) are both smaller than that in Fig. 3(d), while at a higher pulse energy Fig. 3(f) shows the larger spectral width and redshift compared to Fig. 3(d). In addition, in Figs. 3(d)-3(f), the phase-matched DWs are both generated in the region of $\beta_A > 0$ where the Kerr-induced contribution is always greater than the contribution of plasma, and the $\beta_B^{\max}$ is fixed at the similar position since the Kerr-induced contributions show the similar values for different input pulse energies in the absence of plasma.

## 4. CONCLUSIONS

In conclusion, we experimentally and theoretically demonstrated that the plasma-driven blueshifting soliton can strongly enhance the emission of resonance-induced phase-matched DW in a 25-cm-long He-filled SR-PCF. At the input pulse energy of 6 μJ, we observed that the generated DW shows a notable spectral ratio of ~53% and a high energy conversion efficiency of ~19%. In addition, the spectral bandwidth of the DW becomes wider as input pulse energy increases, and its spectrum broadens to longer wavelengths, showing a redshift. This is the direct consequence of phase-matched coupling between the plasma-driven blueshifting soliton and the DW. These experimental results, well-supported by numerical simulations, offer some useful insight in understanding the resonance-induced phase-matched DW emission in gas-filled HC-PCFs. Moreover, the set-up demonstrated here provides a simple and efficient light source at visible wavelengths, which could be easily extended in the future to other wavelengths (UV for example) through using properly-designed anti-resonant fibers with resonance bands at different wavelengths.

**Funding.** Program of Shanghai Academic/Technology Research Leader (18XD1404200); National Natural Science Foundation of China (61925507); Strategic Priority Research Program of the Chinese Academy of Sciences (CAS) (XDB16030100); Major Project Science and Technology Commission of Shanghai Municipal (STCSM) (2017SHZDZX02).